\documentclass[12pt]{iopart}

\usepackage{graphicx,iopams,ulem,color}

\def\In{$^{115}$In}
\def\PuCoIn{PuCoIn$_5$}
\def\PuIn{PuIn$_3$}
\def\PuCoGa{PuCoGa$_5$}
\def\PuRhGa{PuRhGa$_5$}

\def\T1{$T_1$}
\def\iT1{$T_1^{-1}$}
\def\iTT{$\left(T_1T\right)^{-1}$}
\def\Tc{$T_c$}
\def\nuq{$\nu_Q$}
\def\ita{$\eta$}
\def\f{{\it f}}

\begin{document}

\title{Microscopic Properties of the Heavy-Fermion Superconductor PuCoIn$_5$ Explored by Nuclear Quadrupole Resonance}

\author{G. Koutroulakis}
\address{Los Alamos National Laboratory, Los Alamos, NM 87545}
\ead{gkoutrou@lanl.gov}
\author{H. Yasuoka}
\address{Los Alamos National Laboratory, Los Alamos, NM 87545}
\address{Advanced Science Research Center, Japan Atomic Energy Agency, 
Tokai, Ibaraki, 319-1195, JAPAN}
\author{H. Chudo}
\address{Los Alamos National Laboratory, Los Alamos, NM 87545}
\address{Advanced Science Research Center, Japan Atomic Energy Agency, 
Tokai, Ibaraki, 319-1195, JAPAN}
\author{P. H. Tobash}
\address{Los Alamos National Laboratory, Los Alamos, NM 87545}
\author{J. N. Mitchell}
\address{Los Alamos National Laboratory, Los Alamos, NM 87545}
\author{E. D. Bauer}
\address{Los Alamos National Laboratory, Los Alamos, NM 87545}
\author{J. D. Thompson}
\address{Los Alamos National Laboratory, Los Alamos, NM 87545}

\begin{abstract}
We report \In\ nuclear quadrupolar resonance (NQR) measurements on the heavy-fermion superconductor \PuCoIn, in the temperature range \mbox{$0.29{\rm K}\leq T\leq 75{\rm K}$}. The NQR parameters for the two crystallographically inequivalent In sites are determined, and their temperature dependence is investigated. A linear shift of the quadrupolar frequency with lowering temperature below the critical value \Tc\ is revealed, in agreement with the prediction for {\it composite pairing}. The nuclear spin-lattice relaxation rate $T_1^{-1}(T)$ clearly signals a superconducting (SC) phase transition at $T_c\simeq 2.3$K, with strong spin fluctuations, mostly in-plane, dominating the relaxation process in the normal state near to \Tc . Analysis of the \iT1\ data in the SC state suggests that \PuCoIn\ is a strong-coupling $d$-wave superconductor.

\end{abstract}


\maketitle
\section{Introduction}

The character of \f\ electrons in rare earth and actinide materials, i.e. itinerant vs localized, has been the subject of considerable research effort, yet a complete description is still missing. Of particular interest is the case of plutonium, which bridges the itinerant behavior in lighter actinides where 5\f\ electrons form fairly broad conduction bands and the localized, atomic-like 5\f\ states in heavier actinides \cite{Johansson:1975br}. This somewhat dual nature of Pu's 5\f\ states is linked to the variety of unusual properties displayed by elemental Pu and its compounds, as manifested, for example, in the emergence of exotic magnetism and unconventional superconductivity. Characteristically, the Pu-based `115' heavy-fermion materials exhibit a superconducting (SC) transition at critical temperature $T_c=18.5$K, 8.7K for \PuCoGa, \PuRhGa\ respectively \cite{Sarrao:2002jw,Wastin:2003wh}, an order of magnitude higher than any other Ce- or U- based superconductor. Even though the unconventional character of superconductivity in these compounds is widely accepted, the origin of the relatively high \Tc\ and, accordingly, the microscopic mechanism providing the glue for the SC condensate lack an unambiguous explanation. The most prominent relevant ideas propose antiferromagnetic (AF) spin fluctuations  \cite{Monthoux:2001cs}, valence fluctuations \cite{Miyake:2002jd}, or more complex mechanisms \cite{Flint:2008fk} as responsible for mediating superconductivity.

Recently, \PuCoIn, the first In analog of the Pu `115' family, was synthesized, with its physical properties classifying it as a moderately heavy-fermion compound \cite{Bauer:2011jl}. \PuCoIn\ shows a SC transition at \Tc=2.5K, a much lower value than the ones measured for its Ga counterparts. The unit cell of \PuCoIn\ is about 30\% larger than that of \PuCoGa\ and this volume expansion is expected to result in relatively more localized 5\f\ electron states in the former, which has indeed been verified by electronic structure calculations \cite{Zhu:2012bp,Shick:2013fg}. The qualitative difference of the 5\f\ character in these compounds, along with the fact that \PuCoGa\ does not seem to be close to a magnetic instability \cite{Boulet:2005ew}, have put forward the possibility that superconductivity may not be mediated by AF fluctuations in all members of this class of materials as suggested before \cite{Curro:2005fw,Tanaka:2004cu,Pfleiderer:2009gc}. Instead, it is plausible that the high SC transition temperature of \PuCoGa\ is associated with the proximity to a valence instability \cite{Miyake:2007gn}, while superconductivity in \PuCoIn\ is mediated by antiferromagnetic spin fluctuations associated with a quantum critical point \cite{Bauer:2011jl,Shick:2013fg}. An alternative theory that aspires to provide a `universal' solution proposes the development of {\it composite pairs} between local moments and conduction electrons \cite{Flint:2008fk}, and incorporates spin \cite{Flint:2010hc} and valence \cite{Flint:2011cl} fluctuations. Within this picture, both the actinide and the Ce-based heavy-fermion superconductors can be accommodated by appropriately tuning the relative strengths of the model's parameters \cite{Flint:2010hc,Flint:2011cl}.
     
Nuclear quadrupolar resonance (NQR) experiments constitute an ideal tool for the study of \PuCoIn, since they provide a microscopic probe sensitive to both magnetic and charge degrees of freedom. In this report, we present a detailed \In\ NQR investigation of \PuCoIn\ for a wide range of temperature values (0.29K$\leq T\leq$ 75K). From the identified spectral lines, we deduce the quadrupolar parameters for the two inequivalent In sites, which are found to be qualitatively similar to those for other Ce- and Pu-115s. The quadrupolar frequency $\nu_Q$ varies with temperature in the normal state as per the empirical formula for conventional metals. As superconductivity develops, however, \nuq\ exhibits a sharp, albeit small shift, which is a key prediction of the theory of composite SC pairing \cite{Flint:2010hc,Flint:2011cl}. Moreover, the temperature variation of the nuclear spin-lattice relaxation rate \iT1\ delineates distinctive regimes of dynamic behavior. An excess of strong in-plane AF spin fluctuations is observed in the vicinity of $T_c\simeq$2.3K, which are believed to be important for the formation of the SC condensate in this material. Our data below \Tc\ suggest that \PuCoIn\ is a strong-coupling superconductor with anisotropic gap symmetry.

\section{Sample and Experimental details}

Our sample consisted of about 100 mg of \PuCoIn\ single crystals grown from In flux \cite{Bauer:2011jl}, which were ground to powder and placed in a cylindrical coil of 3mm diameter and 7mm length. Prior to sample insertion, in order to prevent any radioactive contamination, the coil was encapsulated in a Stycast 1266 epoxy cast of dimensions $\sim$20mm$\times$20mm$\times$20mm, which was drilled along the coil's axis. Upon sample insertion, the cast's ends were sealed by titanium frits with 2$\mu$m diameter pores, allowing for thermal contact with the respective cooling fluid. For $T\geq1.45$K, temperature was regulated in a standard gas-flow $^4$He cryostat with variations limited to $\delta T/T<1\%$, while, for $T<1.3$K, the coil was mounted into the mixing chamber of a $^3$He/$^4$He dilution refrigerator. Due to the sample's inherent radiation heating, the lowest achieved sample temperature was limited to $T\sim290$mK. The sample temperature was verified by {\it in-situ} measurements of the $^{63}$Cu NMR $T_1$ on the coil's copper nuclei, which should satisfy the relation $T_1T=1.26$sK \cite{NMRshifts}.

The reported NQR spectra were recorded using a conventional pulsed NMR spectrometer. They were obtained by summing the Fourier transforms of standard Hahn spin-echo transients from the $^{115}$In nuclear spins ($I=9/2$), recorded at constant frequency intervals. \T1\ was measured using the \textit{inversion recovery} method: The values were determined by fitting the appropriate function to the magnetization recovery profile after an inversion pulse, depending on the probed nuclear transition.

\section{NQR parameters and discussion}

For nuclei carrying spin $I>1/2$, the nuclear quadrupolar moment $Q$ couples to the local electric field gradient (EFG) created by their surrounding charge distribution. This interaction is described by the Hamiltonian 
\begin{equation}
\mathcal{H_Q}=\frac{h \nu_Q}{6} \left[ 3 \hat{I}_z^2 -  \hat{I}^2 + \frac{1}{2}\eta(\hat{I}_{+}^2 + \hat{I}_{-}^2) \right],
\label{NQR}
\end{equation}
where the characteristic frequency  $\nu_Q$ is defined as $\nu_Q \equiv 3eQV_{ZZ}/(h 2I (2 I -1) )$, $\eta \equiv |V_{XX} -V_{YY}|/|V_{ZZ}|$ is the asymmetry parameter, and $h$ is Planck's constant. Here, $V_{ij} = \partial^2 V/\partial x_i \partial x_j$ are the components of the EFG tensor with the axes labeled according to the convention $|V_{ZZ}|\geq |V_{XX}| \geq |V_{YY}|$, and $e$ is the electron charge. Thus, $\mathcal{H_Q}$ can be fully characterized by $I$, $\nu_Q$, $\eta$, and the unit vector $\hat{n}$ defining the direction of $V_{ZZ}$, the EFG tensor principal axis with the largest eigenvalue.

For $I=9/2$, the eigenstates of eq. \ref{NQR} result in four distinct spectral lines. If the nuclear site possesses uniaxial symmetry, i.e. $\eta=0$, the resonance frequencies are given by integer multiples of $\nu_Q$ ($\nu_i=n\nu_Q$, with $n=1-4$). In case of lower symmetry,  $\eta$ is non-zero and the spectral lines are not equally spaced. 

\PuCoIn\ crystallizes in a tetragonal HoCoGa$_5$-like structure with space group P4/mmm (inset of fig. \ref{fig:nuq_sc}), featuring two crystallographically inequivalent In sites: The high symmetry In(1) site sits in the middle of the basal plane, which corresponds to $\eta=0$ and $\hat{n}\parallel \hat{c}$, hence four equidistant NQR lines should be observed. On the other hand, the lower symmetry In(2) site has orthorhombic symmetry with $\hat{n}$ pointing perpendicular to the face of the unit cell (i.e. $\hat{n}\parallel \hat{a}$ or $\hat{b}$) and $\eta\neq 0$, leading to four not equally separated quadrupolar transitions.

\begin{table}[h]
\caption{Quadrupolar parameters for \PuCoIn\ at $T=3.95$K. The frequencies $\nu_i$ denote the measured spectral lines arising from transitions between the energy levels of Hamiltonian (\ref{NQR}) with $\left|\Delta I_z\right|=1$, so that $\nu_i:\ \left<\left|1/2+i\right|\leftrightarrow \left|1/2+i-1\right|\right>$. The values of $\nu_Q$ and $\eta$ were derived from a $\chi^2$ fit to the data. }
\begin{indented}
\item[]\begin{tabular}{@{}lllllll}
\br
site & $\nu_Q$(MHz) & $\eta$ & $\nu_1$(MHz) & $\nu_2$(MHz) & $\nu_3$(MHz) & $\nu_4$(MHz) \\
\mr
In(1) & 9.434 & 0& - & - & 28.301 & 37.738 \\
In(2) & 15.716 & 0.366& 28.358 & 29.116 & 45.901& 62.180 \\
\br
\end{tabular}
\end{indented}
\label{table:NQRpar_Co}
\end{table}

The \In\ NQR signal in \PuCoIn\ was searched between 20MHz and 90MHz at temperature $T=3.95$K. The detected resonance frequencies are listed in table \ref{table:NQRpar_Co}. The quadrupolar parameters \nuq\ and \ita\  for both \In\ sites were deduced from a $\chi^2$ minimization of the difference between the observed resonance frequencies and the eigenvalues of $\mathcal{H_Q}$, taken from the full diagonalization of eq. \ref{NQR}. The derived  values are $\nu_Q=9.434$MHz and $\eta=0$ for In(1), and $\nu_Q=15.716$MHz and $\eta=0.366$ for In(2). These values are in reasonable agreement with those calculated using the full potential linear-augmented-plane-wave (FLAPW) method within a local density approximation (LDA)  $\nu_Q=11.50$MHz, $\eta=0$, and $\nu_Q=15.97$MHz, $\eta=0.27$ for In(1), and In(2) respectively \cite{Harima}, further attesting to the correct assignment of the spectral lines.

It is important to note that additional spectral lines were detected in the aforementioned frequency range. Upon careful inspection of their temperature dependence, these lines were associated with the quadrupolar transitions of the uniaxially symmetric \In\ nuclear site in a secondary, impurity \PuIn\ phase present in our sample ($\sim17\%$ of sample's mass), with $\nu_Q=10.56$MHz and $\eta=0$. This was verified by independent measurements on a pure \PuIn\ single crystal, which reveal a phase transition to an antiferromagnetically ordered state at $T_N\simeq14$K \cite{PuIn3}. In fact, this impurity phase of \PuIn\ in the \PuCoIn\ sample is responsible for the apparent anomaly at $T\sim14$K in the latter's specific heat \cite{Bauer:2011jl}.
\begin{figure}[h]
\begin{center}
\includegraphics[width=3.3 in]{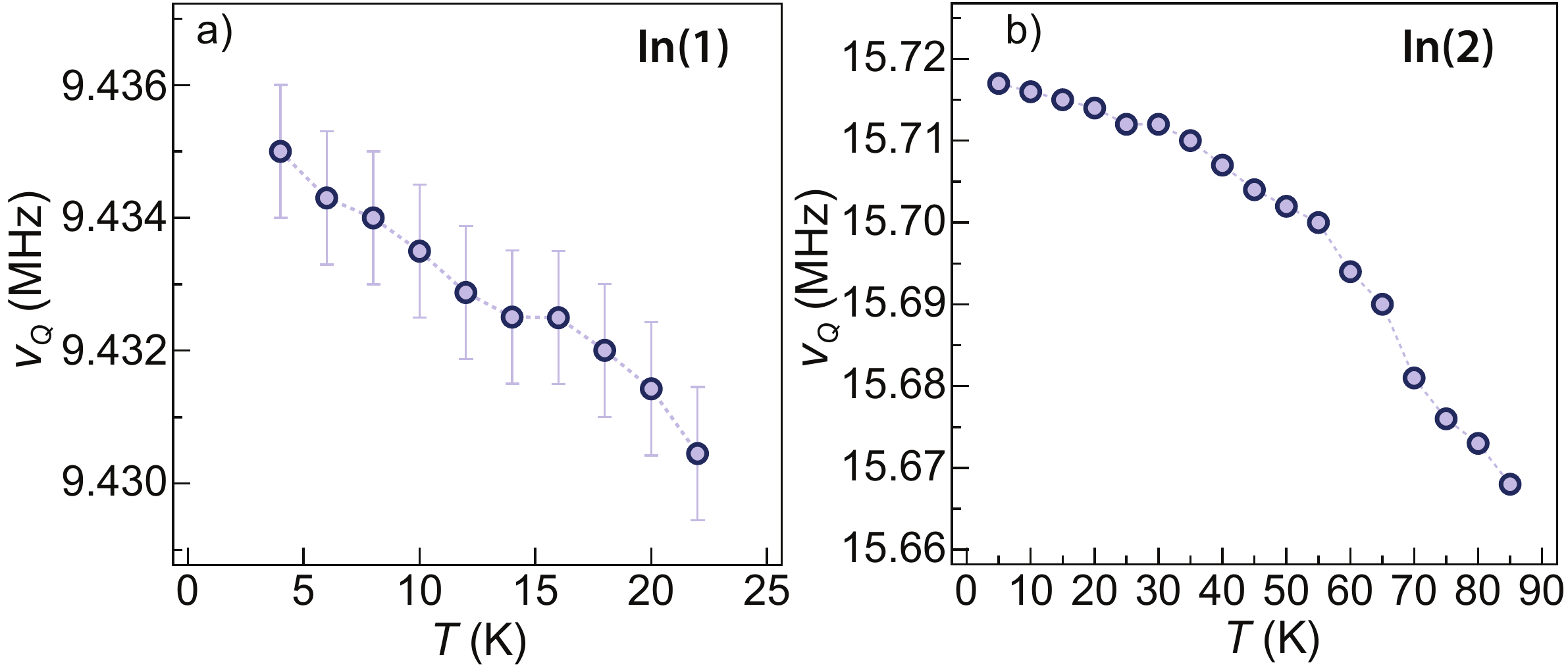}
\caption{Quadrupolar frequency as a function of temperature in the normal state for (a) In(1), and (b) In(2) sites, as discussed in the text.}
\label{fig:nuq}
\end{center}
\end{figure}

The evolution of the quadrupolar frequency as a function of temperature for both \In\ sites, In(1) and In(2), is depicted in Fig.\ref{fig:nuq}a and Fig.\ref{fig:nuq}b, respectively, for $T>T_c$. For In(1), $\nu_Q$ is derived from the position of the $\nu_4\equiv 4\nu_Q$ line, while, for In(2), $\nu_Q$ and $\eta$ at each temperature were deduced using a fit of the measured frequencies $\nu_3$, $\nu_4$ to the eigenvalues of eq. \ref{NQR}, as described above. The quadrupolar frequency increases with decreasing temperature for both nuclear sites, while $\eta$ is only very weakly affected. In general, in conventional non-cubic metals, the temperature dependence of $\nu_Q$ due to the lattice contraction can be described by the empirical relation 
\begin{equation}
\nu_Q(T)=\nu_Q(0)\left(1-A\cdot T^{3/2}\right), \ A>0,
\label{eq:emp}
\end{equation}
where there exists a correlation between the magnitude of $\nu_Q(0)$ and the strength of the EFG's temperature variation, quantified by the coefficient $A$  \cite{Christiansen:1976wh}. Specifically, a larger value of $\nu_Q(0)$, arising from a stronger coupling of the surrounding non-spherical  charge clouds, results in a smaller value of $A$. A least-squares fit of our data to eq.2 yields the following values: $\nu_Q(0)=9.435$MHz, $A=4.44\cdot10^{-6}$K$^{-3/2}$ for In(1), and $\nu_Q(0)=15.718$MHz, $A=3.01\cdot10^{-6}$K$^{-3/2}$ for In(2). Based on this result, the fractional change of $\nu_Q$ as a function of temperature is plotted in fig.\ref{fig:nuq_sc}. It is evident that a $T^{3/2}$ behavior describes well the data for both sites. Moreover, the value of $A$ is indeed lower for In(2), the site with the higher $\nu_Q(0)$, as indicated by the slope of the fit curves in fig.\ref{fig:nuq_sc}. Thus, the effect of temperature on the \In\ quadrupolar parameterds in \PuCoIn\ in the normal state seems to conform to the phenomenological description set out by eq.\ref{eq:emp}.	
\begin{figure}[h]
\begin{center}
\includegraphics[width=3 in]{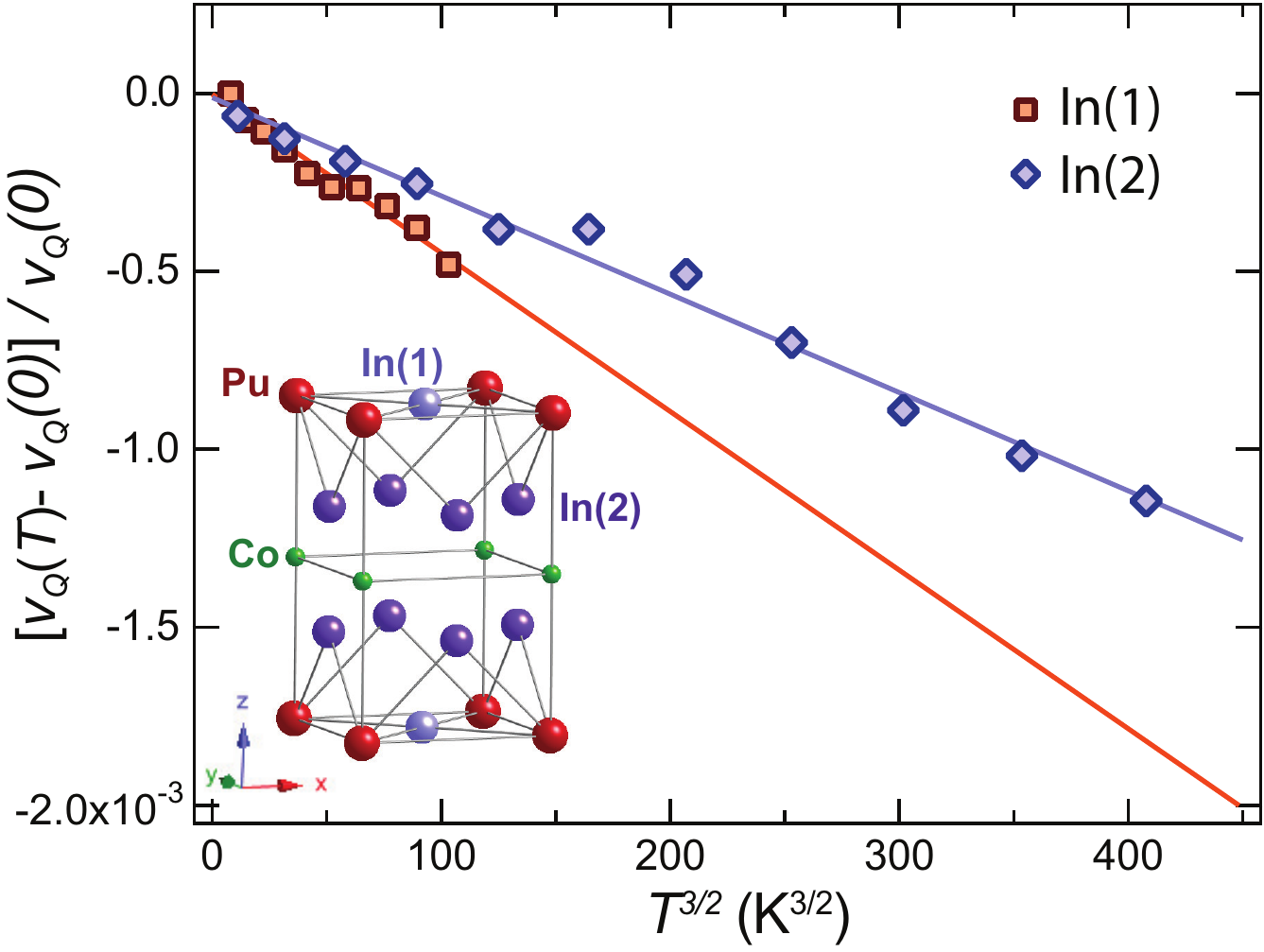}
\caption{Temperature dependence of the fractional change of $\nu_Q$ in the normal state for the two \In\ sites. {\it Inset}: Crystal unit cell structure of \PuCoIn.}
\label{fig:nuq_sc}
\end{center}
\end{figure}

\begin{figure}[h]
\begin{center}
\includegraphics[width=3.2 in]{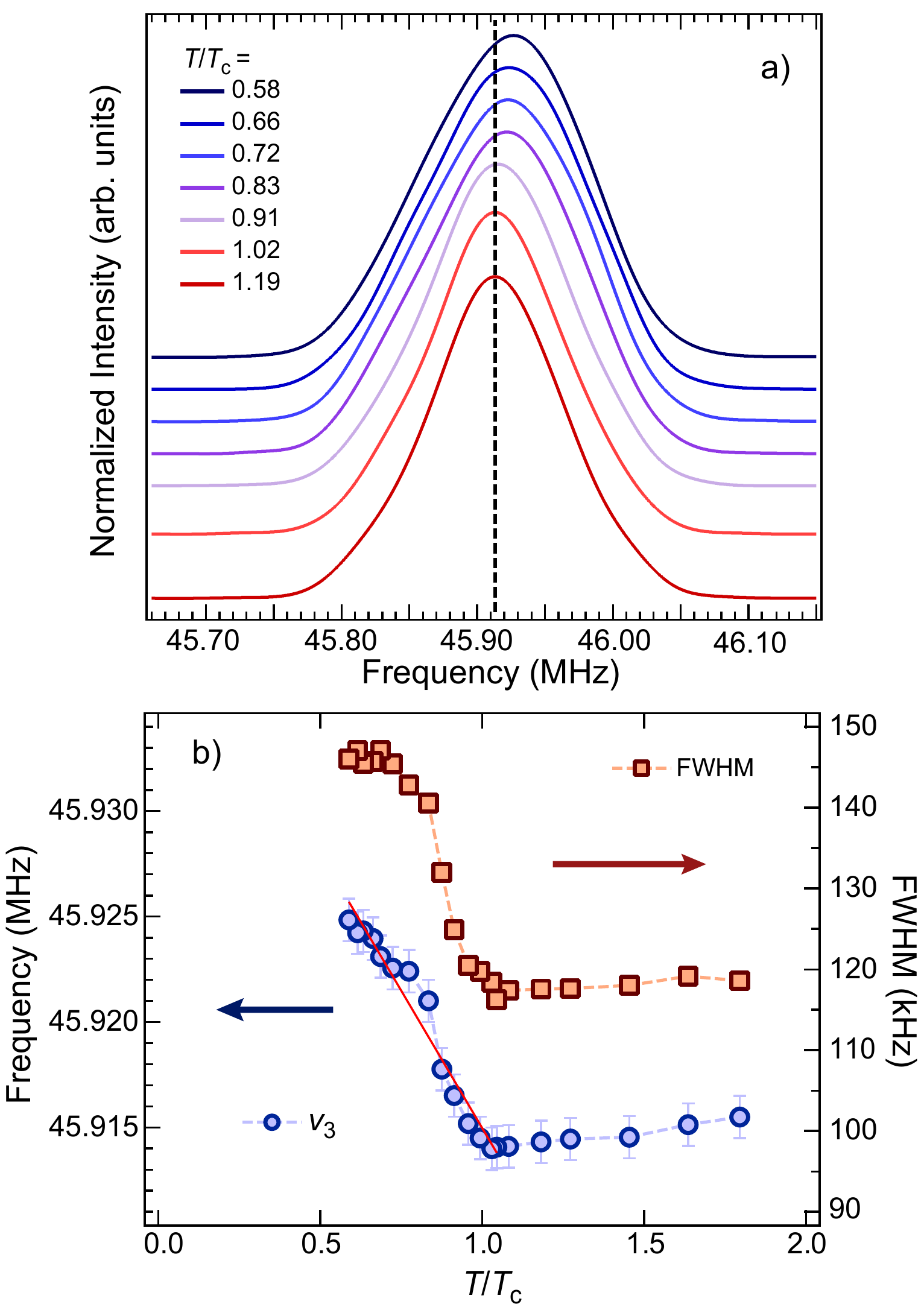}
\caption{a) NQR spectra for the $\left<\pm 7/2\leftrightarrow \pm5/2\right>$ transition of the In(2) site at different values of scaled temperature $T/T_c$, as denoted by the legend. The dashed line marks the relevant resonance frequency just above \Tc. b) First moment (left) and full-width-at-half-maximum (right) of the spectrum described in (a) as a function of temperature near \Tc. The red solid line depicts a linear fit to the data in the SC state.}
\label{fig:nearTc}
\end{center}
\end{figure}

The transition line $\nu_3$ for the In(2) site was carefully traced as a function of temperature in the vicinity of $T_c$, and representative spectra are shown in Fig. \ref{fig:nearTc}a. The resonance frequency deduced from the spectrum's {\it first moment}, and the line's full-width-at-half-maximum (FWHM) are plotted in Fig. \ref{fig:nearTc}b. Even though the position change is small relative to the linewidth, a sharp positive frequency shift is evident in the SC state, beginning precisely at $T_c$ \footnote{The exact value of $T_c$ was verified by {\it in-situ} ac-susceptibility and \T1\  measurements.}. In particular, a linear fit to the data (red solid line in Fig. \ref{fig:nearTc}b) gives an increase of $9.4$kHz/K for $\nu_3$ below \Tc, which corresponds to a change of $\sim 3$kHz/K for \nuq. This observation indicates that the Pu 5\f-electron charge degrees of freedom are readily affected by the emergence of the SC condensate, leading to an altered EFG and hence a shift in \nuq. This change in \nuq\ upon entering the superconducting state is not expected within models of uncovnentional superconductivity mediated by antiferromagnetic spin fluctuations \cite{Monthoux:2001cs} or valence fluctuations \cite{Miyake:2002jd},\cite{Miyake:2007gn}, which predicts a smooth variation of the valence in the superconducting state (see Fig. 6 in ref. \cite{Miyake:2007gn}).  However, such a shift of \nuq\ {\it{is}} predicted to occur within the theoretical framework of {\it composite pairing} \cite{Flint:2010hc,Flint:2011cl}. Specifically, the composite pair condensate is electrostatically active, as its formation results in the redistribution of the \f-electron charge and thus the change of the EFG around the nucleus, which in turn gives rise to a shift in \nuq, contrary to the case of conventional superconductivity. Although the details of the effect's manifestation strongly depend on the symmetry of the model's two distinct scattering channels in the particular material, and their relative strengths \cite{Flint:2010hc,Flint:2011cl}, our present finding in \PuCoIn\ provides evidence for possible composite pair mediated superconductivity. 

It is worth noting that a variation in the electronic density of states arising from the thermal expansion below the SC transition results in an EFG change too small to account for the observed \nuq\ shift. However, changes in \nuq\ in the SC state have been previously observed in some conventional superconductors such as In \cite{Simmons:1961cb} and Ga \cite{Hammond:1960he}, as well as in the high-$T_c$ Ba$_2$YCu$_3$O$_7$ \cite{Riesemeier:1987kw}, warranting caution in the phenomenon's interpretation. Hence, further studies and verification of this small, but discernible shift in \nuq\ below \Tc\ in other members of the \mbox{Pu$MX_5$} family and/or in Ce-115s are necessary to validate the role of composite pairing in heavy-fermion superconductivity.

\section{\T1 measurements and discussion}

The nuclear spin-lattice relaxation rate \iT1\  characterizes the time scale in which the nuclear ensemble reaches thermal equilibrium. This mechanism originates in the coupling of the nuclei to the fluctuations of the local field created by their surrounding {\it lattice}. \iT1\ is directly related to the dynamical spin susceptibility $\chi''(\mathbf{q},\omega_0)$ as \cite{Moriya:1963hp}:
\begin{equation}
\left(\frac{1}{T_1T}\right)_{\parallel}  \propto\sum_{\mathbf{q}}\left[\gamma_n A_{\perp}(\mathbf{q})\right]^2\frac{\chi_{\perp}''(\mathbf{q},\omega_0)}{\omega_0} ,
\label{eq:moriya}
\end{equation}
where $\gamma_n$ is the nucleus' gyromagnetic ratio, $A(\mathbf{q})$ is the $\mathbf{q}$-dependent hyperfine coupling constant, $\omega_0$ is the Larmor frequency, and $\parallel$ ($\perp$) denotes the direction parallel (perpendicular) to the quantization axis of the nuclear spins. Hence, $T_1$ measurements provide an excellent probe of the system's spin dynamics.

\begin{figure}[h]
\begin{center}
\includegraphics[width=3.3 in]{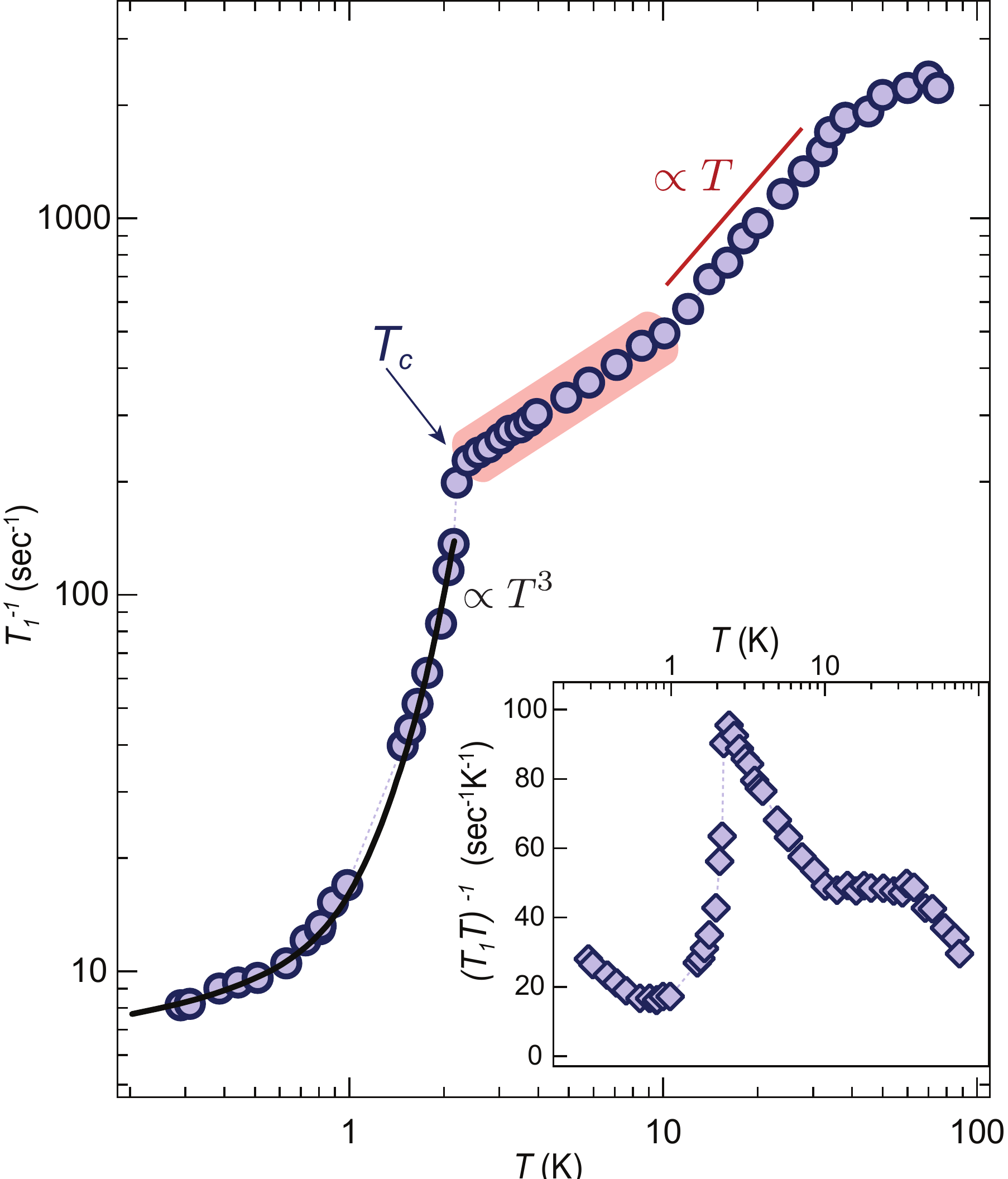}
\caption{\In\ NQR spin-lattice relaxation rate {\it vs.} temperature for the In(2) site. The black solid line depicts the calculation for the SC state relaxation described in the text. The shaded region highlights the temperature range marked by strong AF spin fluctuations. Error bars are within the symbol size. {Inset}: \iTT\ vs. temperature for the In(2) site.}
\label{fig:invT1}
\end{center}
\end{figure}
Figure \ref{fig:invT1} shows the NQR relaxation rate \iT1\ as a function of temperature for $T=$0.29K-75K. The measurements were performed on the $\nu_3$ line, $\left<\pm 7/2\leftrightarrow\pm 5/2\right>$, of the In(2) site (see table \ref{table:NQRpar_Co}). Above \mbox{$T\sim$60K}, \iT1\ is nearly temperature independent, indicating that, in this temperature range, the Pu 5$f$ electronic spins act like localized moments, uncorrelated with the conduction electrons. The In nuclei see the fluctuating local field from these exchange-coupled 5$f$-electrons via a transferred hyperfine interaction, giving rise to the temperature independent relaxation. The absence of Kondo coherence in this temperature region is corroborated by the electrical resistivity behavior, which displays a weak temperature dependence above \mbox{$T\sim$70K}  \cite{Bauer:2011jl}.

As the Pu 5$f$-electrons hybridize with the itinerant electrons and the Kondo lattice develops, the electrical resistivity decreases rapidly with lowering $T$ \cite{Bauer:2011jl}, a hallmark of the emergence of the heavy-fermion state \cite{Fisk:1988fm}. In this regime, the low-lying magnetic excitations are expected to be heavy quasiparticles due to electron-hole pair excitations across the Fermi surface, and this should translate into a {\it Korringa-type} relaxation, i.e. constant \iTT. Indeed, for 50K$\gtrsim T\gtrsim$10K, it is evident that \iT1\ is proportional to $T$, attesting that the system's electronic properties reflect the coherent heavy Fermi-liquid state of the Kondo lattice  in this temperature range.

Below $T\simeq$10K, the relaxation rate deviates markedly from the $T$-linear Korringa behavior down to $T_c\simeq$2.3K \cite{TcNote}, as highlighted by the shaded area in Fig. \ref{fig:invT1}. This trend becomes more evident when plotting \iTT {\it vs.} $T$, shown in the inset of Fig. \ref{fig:invT1}, and it is qualitatively similar to the observations for \PuCoGa\ \cite{Curro:2005fw,Baek:2010ca}. The apparent enhancement of \iTT\ hints at the presence of strong antiferromagnetic spin fluctuations near to \Tc\ \cite{Ishigaki:1996dj}, which may be important for stabilizing the unconventional SC condensate.

The temperature evolution of \iT1\ below \Tc\ demonstrates the non-conventional pairing symmetry in this system: The onset of superconductivity is clearly discerned by the sharp drop of \iT1\ below $T_c\simeq$2.3K. Nevertheless, a coherence peak is not detected in the vicinity of \Tc\, and \iT1 follows a $T^3$ power-law as $T$ decreases just below \Tc. Both these observations contradict the expectations for an isotropic SC gap with conventional $s$-wave symmetry. In contrast, the observed \iT1\ behavior can be replicated successfully assuming an anisotropic line-nodal gap, as illustrated by the solid black line in Fig. \ref{fig:invT1}. Specifically, the temperature dependence of the relaxation rate in the SC state is given by the following equation:
\begin{equation}
\frac{1}{T_{1}^{SC}(T)} = \frac{2}{k_{\rm B}T}\int \Bigl\langle  \frac{N_{\rm s}^2(E)}{N_{0}^{2}} \Bigr\rangle f(E)[1-f(E)] dE.
\label{eq:scT1}
\end{equation}
Here, $f(E)$ is the Fermi-Dirac distribution function, $N_0$ is the normal state density of states (DOS), $N_{\rm s}(E)=N_{0} E/\sqrt{E^{2}-\Delta^{2}(\theta,\phi)}$ is the DOS in the SC state, and $\langle\cdots\rangle$ denotes an average over the Fermi surface. The anisotropic line-nodal gap is taken to be $\Delta(\theta,\phi)\equiv \Delta_0(T)\cos\theta$, where $\theta$ and $\phi$ define the Fermi surface angular parameters and $\Delta_0(T)$ is the BCS gap function. Typically, at low temperature and well below $T_c$, \iT1\ in the d-wave SC state deviates from the $T^3$ behavior due to impurities which contribute a residual DOS $N_{\rm res}$ at the Fermi level, leading to a $T$-linear temperature dependence. Nevertheless, in our case, this behavior of \iT1\ is further masked by the large impurity scattering associated with crystal defects caused by the radioactive decay of Pu and secondary phases. Consequently, we take into consideration an impurity relaxation term in order to account for our data. Accordingly, the total low-temperature relaxation rate in the SC state is calculated as  
\begin{equation}
1/T_1(T)\equiv 1/T_1^{imp}+[ 1/T_1(T_c)-1/T_1^{imp}]\cdot 1/T_1^{SC}(T), 
\label{eq:scT1_imp}
\end{equation}
where $1/T_1^{imp}$ is the impurity contribution and $1/T_{1}^{SC}$ given by Eq. \ref{eq:scT1}. 

A calculation of \iT1 per Eq. \ref{eq:scT1_imp}, for a SC gap $2\Delta_0(0)/k_{\rm B}T_c=8$ with a residual DOS $N_{res}/N_0\simeq 0.32$ and $1/T_1^{imp}$=6.5s$^{-1}$, is illustrated by the solid black line in Fig. \ref{fig:invT1}. The amplitude of the gap is very similar to that of \PuCoGa, $2\Delta_0/k_{\rm B}T_c=6.4-8$ \cite{Daghero:2012fe,Curro:2005fw}, larger than that of \PuRhGa, $2\Delta_0/k_{\rm B}T_c\simeq 5$ \cite{Sakai:2005fh}, and also much larger that the value of 4.28 predicted for a weak electron-boson coupling $d$-wave superconductor. In principle, it is not possible to distinguish with certainty between spin-singlet and -triplet pairing nodal superconductivity solely from the \iT1 temperature dependence and without measurements of the NMR shift. Nevertheless, such measurements in the isostructural \PuCoGa\ \cite{Curro:2005fw,Baek:2010ca} and \PuRhGa\ \cite{Sakai:2006uf}  have provided clear evidence for singlet-pairing, rendering the possibility of p-wave triplet pairing in \PuCoIn\ highly unlikely. Thus, in light of our NQR \iT1\ results, \PuCoIn\ can be classified as an unconventional, strong-coupling $d$-wave superconductor.

\begin{figure}[h]
\begin{center}
\includegraphics[width=3.3 in]{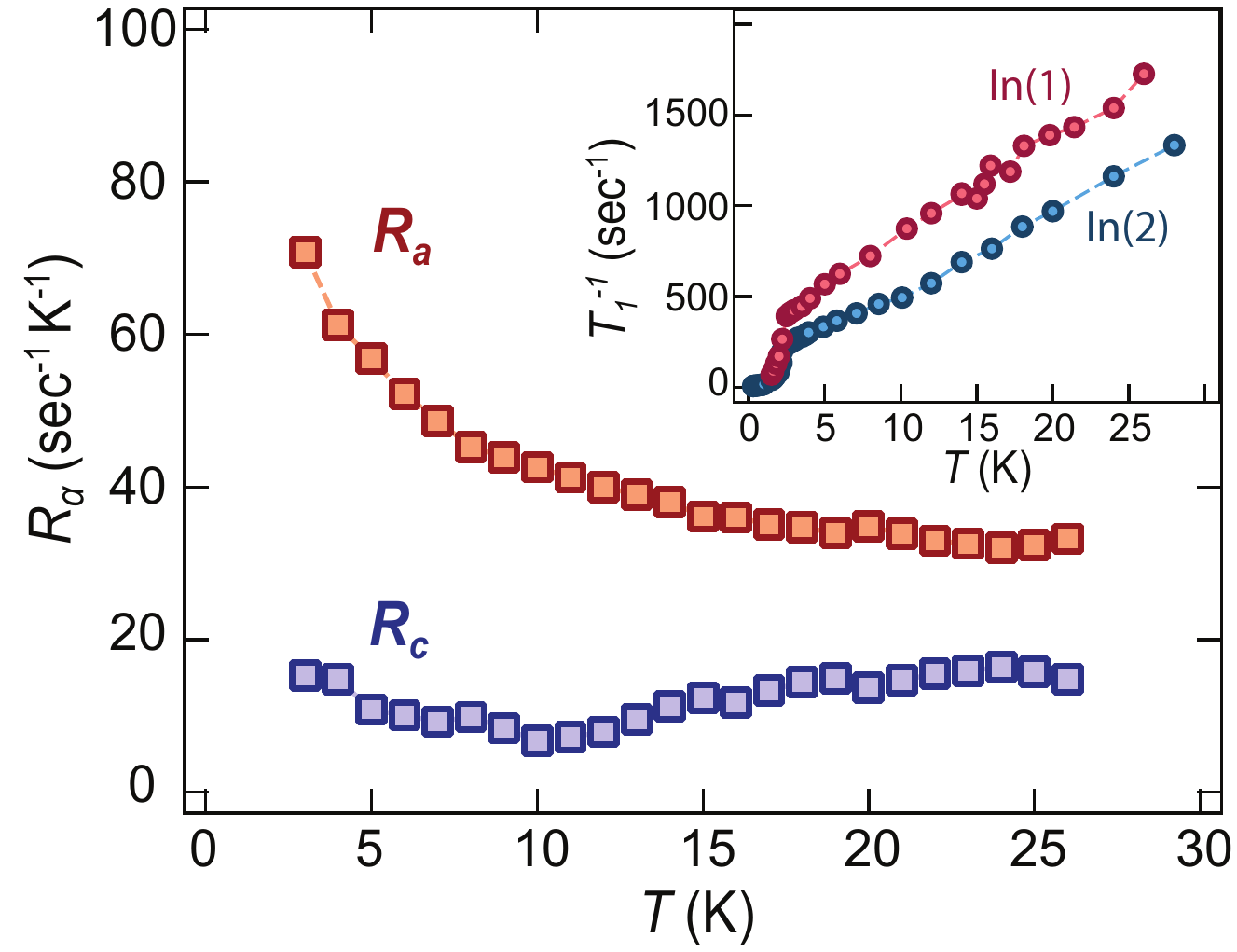}
\caption{Temperature dependence of the rates $R_{\alpha}$ for $\alpha=a,c$, as defined in the text. The {\it inset} shows the $T_1^{-1}(T)$ for both In sites in the normal state, used to derive $R_{\alpha}(T)$.}
\label{fig:Rac}
\end{center}
\end{figure}
In order to further investigate the character of the strong spin fluctuations in the normal state, we examined separately the relaxation rates that are sensitive to fluctuations along different directions, according to Eq. \ref{eq:moriya}. To this end, the rates $R_{\alpha}\equiv \left(\gamma_n A\right)^2\sum_{\mathbf{q}} \chi_{\alpha}''(\mathbf{q},\omega_0)/\omega_0$ are defined, where $\alpha=a,b,c$ \cite{Kambe:2007jm}. Due to the system's tetragonal symmetry, these are reduced to the in- and out-of- plane components $R_a$ and $R_c$, respectively. Then, it follows from Eq. \ref{eq:moriya} that $R_a=1/2\left(T_1T\right)^{-1}_{\parallel c}$ and $R_c=\left(T_1T\right)^{-1}_{\perp c}-1/2\left(T_1T\right)^{-1}_{\parallel c}$. As discussed above, the principal axis of the EFG for the In(1) site is  $\hat{n}\parallel \hat{c}$, while for In(2) it is  $\hat{n}\perp \hat{c}$. That is to say, the quantization axis of the NQR Hamiltonian (Eq. \ref{NQR}) is $\parallel , \perp \hat{c}$ for In(1), In(2), respectively, and thus it is $\left(T_1T\right)^{-1}_{\parallel c}\equiv \left(T_1T\right)^{-1}_{\rm In(1)}$ and $\left(T_1T\right)^{-1}_{\perp c}\equiv \left(T_1T\right)^{-1}_{\rm In(2)}$ (assuming $A(1)\sim A(2)$). Figure \ref{fig:Rac} plots the temperature evolution of the relaxation rates $R_\alpha$ in the normal state, as calculated from the \iT1\ values of the two distinct In sites (see inset). The in-plane component $R_a$ is larger than its out-of-plane counterpart $R_c$ throughout the examined temperature range, and increases rapidly below $T\sim$10K on approaching $T_c$. In contrast, $R_c$ remains nearly unchanged with temperature. This observation suggests that the excess of spin fluctuations reflected in the system's \iTT\ behavior for $T_c\leq T\lesssim$10K stems predominantly from the in-plane component.

\section{Conclusion}

Our \In\ NQR measurements provide a comprehensive picture for the quadrupolar parameters and the dynamical spin susceptibility, via the nuclear spin-lattice relaxation probe, in \PuCoIn, both in the normal and superconducting states. While the quadrupolar frequency temperature variation is typical of a conventional metal above \Tc, the observation of a sharp frequency shift emerging precisely with the development of superconductivity confirms a major prediction of the composite pairing theory. The \iT1\ results reveal the deviation from Fermi-liquid behavior below $T\sim4T_c$ and approaching the superconducting transition, characterized by the appearance of strong in-plane spin fluctuations. Furthermore, below \Tc, the data are accurately reproduced by a line-nodal order parameter calculation with strong coupling. Thus, we conclude that \PuCoIn\ is an anisotropic d-wave superconductor, likely near a magnetic instability, where spin fluctuations appear to be playing a central role in promoting superconductivity.
 
\section{Acknowledgements}

We thank S.E. Brown for his insightful input, and R. Flint, P. Coleman and H. Harima for helpful discussions. Work at Los Alamos National Laboratory was performed under the auspices of the US Department of Energy, Office of Basic Energy Sciences, Division of Materials Sciences and Engineering, and from the Los Alamos Laboratory Directed Research and Development program. G.K. and H.Y. acknowledge support from the Glenn T. Seaborg Institute.
\\

\end{document}